\documentclass[twocolumn]{jpsj3}
\usepackage{txfonts}

\title{
Spin-Hall Effect and Diamagnetism of Anisotropic Dirac Electrons in Solids
}

\author{Yuki Fuseya$^1$\thanks{E-mail: fuseya@pc.uec.ac.jp}, 
Masao Ogata$^2$, 
and 
Hidetoshi Fukuyama$^3$}
\inst{$^1$Department of Engineering Science, University of Electro-Communications, Chofu, Tokyo 182-8585 \\
$^2$Department of Physics, University of Tokyo, 7-3-1 Hongo, Bunkyo-ku, Tokyo 113-0033\\
$^3$Department of Applied Physics and Research Institute for Science and Technology, Tokyo University of Science, Kagurazaka, Shinjuku-ku, Tokyo 162-860} 

\abst{
	Spin-Hall conductivity $\sigma_{{\rm s}xy}$ and orbital susceptibility $\chi$ are investigated for the anisotropic Wolff Hamiltonian, which is an effective Hamiltonian common to Dirac electrons in solids.
	It is found that, both for $\sigma_{{\rm s}xy}$ and $\chi$, the effect of anisotropy appears only in the prefactors, which is given as the Gaussian curvature of the energy dispersion, and their functional forms are equivalent to those of the isotropic Wolff Hamiltonian.
	As a result, it is revealed that the relationship between the spin Hall conductivity and the orbital susceptibility in the insulating state, $\sigma_{{\rm s}xy}=(3mc^2/\hbar e)\chi$, which was firstly derived for the isotropic Wolff Hamiltonian, is also valid for the anisotropic Wolff Hamiltonian.
	Based on this theoretical finding, the magnitude of spin-Hall conductivity is estimated for bismuth and its alloys with antimony by that of orbital susceptibility, which has good correspondence between theory and experiments.
	The magnitude of spin-Hall conductivity turns out to be as large as $e\sigma_{{\rm s}xy} \sim 10^4 {\Omega}^{-1}{\rm cm}^{-1}$, which is about 100 times larger than that of Pt.
}

\kword{Dirac electrons in solids, Wolff Hamiltonian, spin-Hall effect, diamagnetism, bismuth}

\usepackage{amsmath,bm,mathrsfs,color,accents}
\renewcommand{\Im}{{\rm i}}

\newcommand{\D}{\Delta}

\newcommand{\bk}{\bm{k}}
\newcommand{\bp}{\bm{p}}

\newcommand{\br}{\bm{r}}

\newcommand{\bpi}{\bm{\pi}}
\newcommand{\bsgm}{\bm{\sigma}}
\newcommand{\bW}{\bm{W}}
\newcommand{\bgam}{\bm{\gamma}}

\newcommand{\scr}[1]{\mathscr{#1}}
\newcommand{\ve}{\varepsilon}
\newcommand{\tve}{\tilde{\varepsilon}}
\newcommand{\tk}{\tilde{k}}
\newcommand{\sgn}{{\rm sgn}}

\begin{document}
\maketitle

\section{Introduction}
	
	In 1964, Wolff found that the effective Hamiltonian of bismuth is essentially identical to that of the Dirac Hamiltonian.\cite{Wolff1964}
	This finding led to the solution of the long-standing mystery of bismuth, i.e., the large diamagnetism.
	Although bismuth is the first material in which the diamagnetism is verified, the anomalously large diamagnetism cannot be explained by the standard theory of diamagnetism, the Landau-Peierls formula.\cite{Peierls1933}.
	The diamagnetism of bismuth and its alloys with antimony takes its largest value when the chemical potential locates in the band gap\cite{Wehrli1968}.
	On the other hand, the Landau-Peierls formula predicts that the diamagnetism is proportional to the density of states, namely, the diamagnetism should vanish for insulators.
	This mystery was finally solved\cite{Fukuyama1970,Fukuyama2012} by taking into account the interband effect of a magnetic field (not taken into account in the Landau-Peierls formula) and the spin-orbit interaction (quite large for bismuth $\sim 1.5 $eV) based on Wolff's Dirac Hamiltonian.

	Recently, another interesting phenomena associated with bismuth was proposed\cite{Fuseya2012b}: the spin Hall effect (SHE).
	The SHE is an effect that the spin magnetic-moment of electrons flows perpendicular to the external electric field.
	The original mechanisms of SHE proposed by Dyakonov-Perel\cite{Dyakonov1971a,Dyakonov1971b} and Hirsch\cite{Hirsch1999} are basically the effect driven by the scattering due to impurities, so it is called an extrinsic mechanism.
	On the other hand, the possibility of an intrinsic mechanism of SHE was put forward  by Murakami-Nagaosa-Zhang\cite{Murakami2003} and Sinova {\it et al.}\cite{Sinova2004}.
	Unlike the extrinsic mechanism, the intrinsic mechanism is only due to the electronic structure of solids.
	The intrinsic mechanism is possible not only for metals, but also for semiconductors or insulators.
	It is expected that such a spin Hall insulator can generate spin current without any Joule heating, which is highly desired for spintronics.
	However, there are only few cases where the possibility of spin Hall insulator has been examined in actual materials.\cite{Murakami2006}

	For this issue, it was shown that the spin Hall insulator can be achieved in the isotropic Wolff Hamiltonian\cite{Fuseya2012b}, which is a simplified Hamiltonian of original Wolff's Dirac Hamiltonian.
	Furthermore, it was also proved that, when the chemical potential locates in the band gap, the spin Hall conductivity, $\sigma_{{\rm s}xy}$, is related to the orbital susceptibility, $\chi$, through a very simple relationship with only fundamental physical constants as follows\cite{Fuseya2012b,note0}
	\begin{align}
		\sigma_{{\rm s}xy} = \frac{3mc^2}{\hbar e }\chi,
		\label{eq1}
\end{align}
	where $m$ is the bare electron mass, $c$ is the velocity of light, $\hbar$ being the Plank constant divided by $2\pi$ and $e$ ($>0$) the elementary charge.
	(The magnetic moment for the spin-Hall conductivity and the magnetic field for the orbital susceptibility are in the $z$-axis.)
	However, the electronic structure of real materials is anisotropic in general.
	%
	%
	Actually, the electronic structure of bismuth is highly anisotropic.\cite{Smith1964,ZZhu2011,ZZhu2012,Dresselhaus1971,Edelman1976,Issi1979}
	The isotropic Wolff model completely discards these anisotropic characteristic of actual materials.
	Hence it is crucially important to see whether eq. (\ref{eq1}) still holds even in the anisotropic Wolff Hamiltonian in order to be compared with experiments.

	The purpose of the present paper is to calculate the spin Hall conductivity and the orbital susceptibility for anisotropic Wolff Hamiltonian, and to verify eq. (\ref{eq1}) even in this case.
	Based on this finding, we give a quantitative estimation for the magnitude of the spin Hall conductivity. 

	The anisotropic Wolff Hamiltonian for the Dirac electron systems is introduced and basic properties of this Hamiltonian is summarized in \S \ref{Model}.
	In \S\ref{SHC}, the spin Hall conductivity is calculated based on the Kubo formula.
	The orbital susceptibility in the same theoretical framework is calculated in \S\ref{Orbital susceptibility}.
	In \S\ref{Discussion} we discuss implications of the present theoretical results to the experiments.
	\S\ref{Summary} is devoted for the summary.

\section{Anisotropic Wolff Hamiltonian}\label{Model}
	%
	We consider a one electron Hamiltonian with a spin-orbit interaction
	\begin{align}
		\scr{H} = \frac{p^2}{2m}+ V(\br) + \frac{\hbar}{4m^2c^2}\bsgm \cdot \bm{\nabla} V(\br) \times \bp,
		\label{eq2}
\end{align}
	where $V(\br)$ is the periodic crystal potential.
	The $\bk \cdot \bp$ theory for eq. (\ref{eq2}) in the case where only two bands get close in energy will generally be written in terms of $4\times 4$ matrix as follows\cite{Wolff1964}
\begin{align}
	\scr{H}&= \frac{\hbar^2 k^2}{2m}+
	\D \beta + \Im \hbar \bk \cdot \left[ \sum_{\mu=1}^3 \bW(\mu) \beta \alpha_\mu \right]
	\\
	&=
	\begin{pmatrix}
		\D & \Im \hbar \bk\cdot \left[ \sum_{\mu=1}^3 \bW(\mu) \sigma_\mu \right] \\
		-\Im \hbar \bk\cdot \left[ \sum_{\mu=1}^3 \bW(\mu) \sigma_\mu \right] & -\D
	\end{pmatrix},
	\label{Wolff}
\end{align}
	where $\bk $ is measured from an extremum of energy band, the energy gap is $2\D$, and the origin of the energy is taken at the center of the energy gap.
	Equation (\ref{Wolff}) is basically the same as the Dirac Hamiltonian, but spatially anisotropic.
	The $4\times 4$ matrices $\alpha_\mu$ and $\beta$ are the same as that appear in the original Dirac theory
	\begin{align}
		\alpha_\mu = 
		\begin{pmatrix}
			0 & \sigma_\mu \\
			\sigma_\mu & 0
		\end{pmatrix},
		\quad
		\beta =
		\begin{pmatrix}
			I & 0 \\
			0 & -I
		\end{pmatrix}.
\end{align}
	Three real vectors $\bW (\mu)$ ($\mu=1, 2, 3$) represent the matrix elements of the velocity operator,
	\begin{align}
		\bm{v}&= \frac{\bp}{m}+\frac{\hbar}{4m^2 c^2}\bsgm \cdot \bm{\nabla} V(\br),
		\label{eq6}
	\end{align}
	as
	\begin{align}
		\bW (1) &= {\rm Im} \left[ \langle 1 | \bm{v} | 4 \rangle \right]
		=-{\rm Im}\left[ \langle 3 | \bm{v} | 2 \rangle \right],\\
		\bW (2) &= {\rm Re} \left[\langle 1 | \bm{v} | 4 \rangle \right]
		=-{\rm Re}\left[ \langle 3 | \bm{v} | 2 \rangle \right],\\
		\bW (3) &= {\rm Im} \left[ \langle 1 | \bm{v} | 3 \rangle \right]
		={\rm Im} \left[ \langle 4 | \bm{v} | 2 \rangle \right],
	\end{align}
	where $|1\rangle \sim |4\rangle$ denote the four band-edge wave functions.
	Here, the time reversal symmetry and the reflection invariance are incorporated, and the real part of the vector $ \langle 1 | \bm{v} | 3 \rangle = \langle 4 | \bm{v} | 2 \rangle$ is set to be zero with an appropriate choice of the basis functions.\cite{Wolff1964}
	%
	%
	%
	We call this Hamiltonian (\ref{Wolff}) the anisotropic Wolff Hamiltonian in the following.
	The eigen energy of this Hamiltonian is obtained as
\begin{align}
	E_k = \frac{\hbar^2 k^2}{2m }\pm \sqrt{\D^2 + \sum_{\mu=1}^3 \left[ \hbar \bk \cdot \bW(\mu)\right]^2}.
\end{align}

	Wolff introduced this Hamiltonian as an effective Hamiltonian for bismuth.
	However, this form is quite general and actually gives an effective Hamiltonian for two-band systems with a small energy gap and a strong spin-orbit interaction.
	The characteristics of each materials is reflected in $\bW(\mu)$ and $\D$.
	Note that the contribution from the quadratic term $\hbar^2 k^2/2m$ is negligibly small when the band gap is small.
	Hereafter, we shall discard the quadratic term.
	When we assume
	\begin{align}
		\bW(1) &= ( \gamma, 0, 0) \\
		\bW(2) &= (0, \gamma, 0)\\
		\bW(3) &= (0, 0, \gamma),
\end{align}
	we have the isotropic Wolff Hamiltonian (the same as the original Dirac Hamiltonian),\cite{Fuseya2009,Fuseya2012a,Fuseya2012b}
	\begin{align}
		\scr{H}_{\rm iso} =
		\begin{pmatrix}
			\D & \Im \hbar \gamma \bk \cdot \bsgm \\
			-\Im \hbar \gamma \bk \cdot \bsgm & -\D
		\end{pmatrix}.
\end{align}


	In order to investigate the SHE, we need to define the spin magnetic-moment, which is correctly determined by considering the Hamiltonian under a magnetic field as follows.
	In the presence of a magnetic field $\bm{B}$, the Hamiltonian is obtained by the replacement $\hbar \bk \to \bpi = (\hbar \bk + e\bm{A}/c)$ with $\bm{A}$ being the vector potential in the present $\bk \cdot \bp$ theory.
	The corresponding Schr\"odinger equation becomes
\begin{align}
	\scr{H}\psi = 
	\begin{pmatrix}
		\D & \Im \bpi \cdot \bm{\gamma} \\
		-\Im \bpi \cdot \bm{\gamma} & -\D
	\end{pmatrix}
	\psi = E\psi,
\end{align}
	where
\begin{align}
	\bm{\gamma} = \sum_\mu \bW(\mu) \sigma_\mu.
\end{align}
	The energy eigenvalues of this equation is easily obtained by considering its square, $\scr{H}^2 \psi = E^2 \psi$, as
\begin{align}
	\scr{H}^2 &= 
	\begin{pmatrix}
		\D^2 + \left( \bpi \cdot \bm{\gamma} \right)^2 & 0\\
		0 & \D^2 +  \left( \bpi \cdot \bm{\gamma} \right)^2
	\end{pmatrix},
	\nonumber\\
	&=
	\begin{pmatrix}
		\D^2 + 2\D \scr{H}^* & 0\\
		0 & \D^2 + 2\D \scr{H}^*
	\end{pmatrix},
	\\
	\scr{H}^* &= \frac{\bpi \cdot \hat{\alpha}\cdot \bpi}{2}+ \bm{\mu}_{\rm s}^*\cdot \bm{B}.
\end{align}
	where $\hat{\alpha}$ and $\bm{\mu}_{\rm s}^*$ are the inverse mass-tensor and the spin magnetic-moment near the extremum of band, which are given by
\begin{align}
	\alpha_{ij} &= \frac{1}{\D} \sum_\mu W_i (\mu) W_j (\mu),
	\label{inverse alpha}
	\\
	\bm{\mu}_{\rm s}^*&= \frac{\hbar e}{2c}\sum_{\lambda \mu \nu} \epsilon_{\lambda \mu \nu}
	\frac{ \bW(\lambda) \times \bW (\mu) \sigma_\nu }{2\D}
	\nonumber\\
	&=\frac{\hbar e \Omega}{2c\D}
	\left[
	\bm{Q}(1) \sigma_1 + \bm{Q}(2) \sigma_2 + \bm{Q} (3) \sigma_3
	\right] ,
\end{align}
respectively.\cite{Wolff1964}
	Here we noted
	\begin{align}
	\left( \bpi \cdot \bm{\gamma} \right)^2
	&= \sum_\mu \left[ \bpi \cdot \bW(\mu) \right]^2 + 2\D \bm{\mu}_{\rm s}^* \cdot \bm{B},
	\\
	\bm{Q}(\lambda) &= \frac{\bW(\mu) \times \bW(\nu)}{\Omega},
	\quad \text{($\lambda, \mu, \nu$ cyclic)}
	\\
	\Omega &= \bW(1) \times \bW(2) \cdot \bW(3),
\end{align}
	and $\epsilon_{\lambda \mu \nu}$ is Levi-Civita's totally antisymmetric tensor.
	%
	%

	
	By the Foldy-Wouthuysen transformation,\cite{Foldy1950,Schweber_text} the original anisotropic Wolff Hamiltonian is transformed to 
	\begin{align}
		\scr{H}_{\rm FW}=
		\begin{pmatrix}
			\D + \scr{H}^* & 0\\
			0 & -\D -\scr{H}^*
		\end{pmatrix}.
		\label{FW}
\end{align}
	This indicates that the signs of the spin magnetic-moment, which is given by the coefficient of $\bm{B}$, are opposite between the conduction and valence bands.
	Consequently, the spin magnetic moment in a $4\times 4$ matrix is defined by
	\begin{align}
	\bm{\mu}_{\rm s}=
	\begin{pmatrix}
		-\bm{\mu}_{\rm s}^* & 0\\
		0 & \bm{\mu}_{\rm s}^*
	\end{pmatrix}.
\end{align}
	The velocity of the spin magnetic-moment can be defined as a tensor	
\begin{align}
	v_{{\rm s}j}^{i}&=\frac{1}{2\mu_{\rm B}} \left\{ \mu_{{\rm s}i},  v_j  \right\}
	\nonumber\\
	&=\frac{1}{\mu_{\rm B}}\frac{\hbar e \Omega}{2c\D}
	\sum_{\lambda \mu \nu}\epsilon_{\lambda \mu \nu}
	\begin{pmatrix}
		0 & Q_i(\lambda)W_j(\mu) \sigma_\nu \\
		Q_i(\lambda) W_j (\mu) \sigma_\nu & 0
	\end{pmatrix},
	\label{spin_velocity}
\end{align}
	where $\mu_{\rm B}=\hbar e/2mc$ is the Bohr magneton and $v_j$ is the velocity operator given by
	\begin{align}
	\bm{v} = \frac{1}{\hbar }\frac{\partial \scr{H}}{\partial \bk}= 
	\begin{pmatrix}
		0 & \Im \sum_\mu \bW (\mu) \sigma_\mu \\
		-\Im \sum_\mu \bW (\mu) \sigma_\mu & 0
	\end{pmatrix}.
\end{align}
	%

\section{Spin-Hall conductivity}\label{SHC}
	%
	In the present paper, we define the spin-Hall conductivity tensor as a linear response of the ``spin current" to the electric field, and consider the velocity of the spin magnetic-moment defined by eq. (\ref{spin_velocity}) as the spin-current operator.\cite{Fuseya2012a,Fuseya2012b}
	Then the spin-Hall conductivity is given by
	\begin{align}
		\sigma_{{\rm s}jk}^i &= \frac{1}{\Im \omega}
		\left[
		\Phi_{{\rm s}jk}^i(\omega + \Im \delta)-\Phi_{{\rm s}jk}^i (0 + \Im \delta)
		\right],
		\\
		\Phi_{{\rm s}jk}^i(\Im \omega_\lambda)&=
		-e k_{\rm B}T \sum_{n \bk} {\rm Tr}
		\left[
		\scr{G}(\Im \tilde{\ve}_n) v_{{\rm s}j}^i \scr{G}(\Im \tilde{\ve}_{n-}) v_k
		\right],
		\label{eq26}
\end{align}
	where $\ve_n = (2n+1) \pi k_{\rm B}T$, $\ve_{n-}= \ve_n - \hbar \omega_\lambda$, and $\omega_\lambda = 2\pi \lambda k_{\rm B}T$ ($n, \lambda$: integer) to be analytically continued as $\Im \omega_\lambda \to \omega + \Im \delta$ ($\omega$ is the frequency of the external electric field).
	We introduced a quasiparticle damping rate $\Gamma $ by the form $\Im \tilde{\ve}_n = \Im \ve_n + \Im \Gamma \sgn (\ve_n)$.
	The Green's function, $\scr{G}(\Im \ve_n) = [\Im \tilde{\ve}_n - \scr{H}]^{-1}$, is given as
\begin{align}
	\scr{G}(\Im \ve_n) 
	=\frac{1}{(\Im \tilde{\ve}_n)^2 - E_k^2}
	\begin{pmatrix}
		\Im \tilde{\ve}_n + \D & \Im \bk \cdot \bm{\gamma} \\
		-\Im \bk \cdot \bm{\gamma} & \Im \tilde{\ve}_n - \D
	\end{pmatrix}.
\end{align}
	The trace of $\scr{G}v_{{\rm s}j}^i \scr{G}_-v_k$ in eq. (\ref{eq26}) is evaluated to be
\begin{align}
	{\rm Tr} \left[
	\scr{G}v_{{\rm s}j}^i \scr{G}_-v_k
	\right]
	&=
	\frac{4\Im m \Omega  (\Im \tilde{\ve}_n - \Im \tilde{\ve}_{n-})}{\{ (\Im \tilde{\ve}_n)^2 -E_k^2 \}\{ (\Im \tilde{\ve}_{n-})^2 -E_k^2 \}}
	\nonumber\\
	&\times \left[
	\sum_{\lambda \mu \nu} \epsilon_{\lambda \mu \nu}
	Q_i(\lambda)W_j(\mu)W_k(\nu)
	\right]
	.
	\label{eq39}
\end{align}
	(We abbreviate $\scr{G}(\Im \tve_n)$ and $\scr{G}(\Im \tve_{n-})$ as $\scr{G}$ and $\scr{G}_-$, respectively.)
	%
%

\begin{figure}
\begin{center}
\includegraphics[width=70mm]{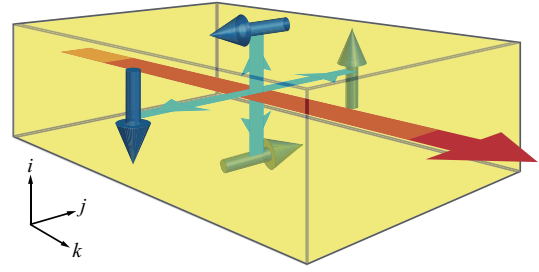}
\end{center}
\caption{(Color online) Illustration of the directions of the electric field (wide arrow), the spin-current (narrow arrows) and the spin magnetic-moment (three-dimensional arrows).}
\label{illust}
\end{figure}

	The spatial anisotropy of the spin-Hall conductivity is represented by the factor $\sum_{\lambda \mu \nu} \epsilon_{\lambda \mu \nu}
	Q_i(\lambda)W_j(\mu)W_k(\nu)$.
	However, the vector $\bW(\mu)$ depends on the representation of the basis and is not observable.
	Instead, as shown in Appendix\ref{Gaussian}, the following relation holds
	\begin{align}
		\sum_{\lambda \mu \nu}\epsilon_{\lambda \mu \nu}Q_i(\lambda) W_j (\mu) W_k(\nu)
		=\frac{\Delta^2}{\Omega} \epsilon_{kji}
		\left(
		\alpha_{jk}^2 - \alpha_{jj}\alpha_{kk}
		\right),
		\label{eq29}
\end{align}
	where $\alpha_{ij}$ is the inverse mass-tensor (\ref{inverse alpha}), which can be determined experimentally\cite{Smith1964,ZZhu2011,ZZhu2012}. 
	The factor $(\alpha_{jk}^2 - \alpha_{jj}\alpha_{kk})$ corresponds to the Gaussian curvature of the energy dispersion.
	It should be emphasized here that the factor $\sum_{\lambda \mu \nu} \epsilon_{\lambda \mu \nu}
	Q_i(\lambda)W_j(\mu)W_k(\nu)$ gives finite contributions only when $i, j, k$ are perpendicular with each other.
	The directions of the electric field, the spin current and the spin magnetic-moment are illustrated in Fig. \ref{illust}.
	Though eq. (\ref{eq29}) can be finite for the other combination, e.g. $Q_i(\lambda)W_j (\mu) W_i (\nu)$, their contributions will vanish in the case with the inversion symmetry, if we take into account all contributions from the whole Brillouin zone.
	The details are given in Appendix\ref{Gaussian}.
	%
	
	%
	For $\bk$-integration in eq. (\ref{eq26}) we introduce a new variable
	\begin{align}
		\tilde{k}_\mu \equiv \bk \cdot \bm{W}(\mu).
\end{align}
	%
	%
	By this transformation, we have 
	\begin{align}
		E&= \sqrt{\D^2 + \sum_\mu [\hbar \bk\cdot \bm{W}(\mu)]^2 } = \sqrt{\D^2 + \hbar^2 \tilde{k}^2},
\end{align}
	with $\tilde{k}^2 = \sum_\mu \tilde{k}_\mu^2$.
	The integration with respect to $\bk$ is transformed as
	\begin{align}
		\int\!\! \frac{d^3 k}{(2\pi)^3} \, g(E_k) =
		\int\!\! \frac{d^3 \tilde{k}}{(2\pi)^3 \sqrt{\D^3 \det \hat{\alpha}}} g(E_{\tilde{k}}),
	\end{align}
	where $g(E)$ is an arbitrary function of $E$. 
	Consequently, the form of $\Phi_{{\rm s}jk}^i (\Im \omega_\lambda)$ becomes essentially equivalent to that for the isotropic case (eq. (7) in Ref. \citen{Fuseya2012b}) except for the Gaussian curvature factor as
	\begin{align}
	\Phi_{{\rm s}jk}^i (\Im \omega_\lambda) &= -e 
	\frac{\epsilon_{kji} \left(
		\alpha_{jk}^2 - \alpha_{jj}\alpha_{kk}
		\right)}
		{\sqrt{\D^3 \det \hat{\alpha}}}
	\nonumber\\
	&\times
	k_{\rm B}T \sum_{n \tilde{k}_\mu }
	\frac{ 4\Im m \D^2( \Im \tilde{\ve}_n - \Im \tilde{\ve}_{n-})}{\{ (\Im \tilde{\ve}_n)^2 -E_{\tilde{k}}^2 \}\{ (\Im \tilde{\ve}_{n-})^2 -E_{\tilde{k}}^2 \}}
	.
	\label{eq34}
\end{align}
	%

\begin{figure}
\begin{center}
\includegraphics[width=80mm]{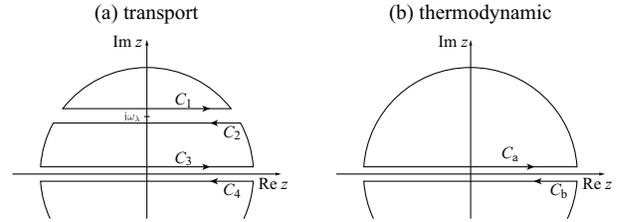}
\end{center}
\caption{Contours of integration for (a) transport coefficient (spin Hall conductivity) and (b) thermodynamic quantity (orbital susceptibility).}
\label{contours}
\end{figure}
	From the standard technique of analytic continuation\cite{AGD_text,Fetter_text} (Appendix\ref{App_Analytic}), we carry out the frequency summation as
	\begin{align}
		k_{\rm B}T\sum_n \mathcal{F}(\Im \ve_n) = -\frac{1}{2\pi \Im} \int_C \!\! dz\, f(z) \mathcal{F}(z),
\end{align}
	where $f(\ve)$ is the Fermi distribution function, and $\mathcal{F}(z)$ is an analytic function in the region enclosed by the contour $C$.
	%
	%
	In the present case, the contour is composed of four contours $C_1 \sim C_4$ dipicted in Fig. \ref{contours} (a).
	After the analytic continuation with respect to $\ve_n$ and then to $\omega_\lambda$, we extract the term linear in $\Im \omega_\lambda \to \omega$ for the dc conductivity.
	Then the contribution from $C_1$ and $C_4$ is
	\begin{align}
	-\frac{\hbar \omega}{2\pi \Im }\int_{-\infty}^{\infty} \!\! d\ve\, f(\ve)
	&
	\left[ 
	\frac{1}{\left( \ve_+^2 -E_{\tilde{k}}^2 \right)^2}
	-\frac{1}
	{\left( \ve_-^2 -E_{\tilde{k}}^2 \right)^2}
	\right],
\end{align}
	and that from $C_2$ and $C_3$ is
	\begin{align}
	-\frac{\hbar \omega}{2\pi \Im } \int_{-\infty}^\infty \!\! d\ve \, 
	\frac{d f(\ve)}{d \ve}
		\frac{2\Im \Gamma}{\left( \ve_+^2 - E_{\tilde{k}}^2 \right)\left( \ve_-^2 - E_{\tilde{k}}^2 \right)},
\end{align}
	where $\ve_\pm = \ve \pm \Im \Gamma$.

\begin{figure}
\begin{center}
\includegraphics[width=7cm]{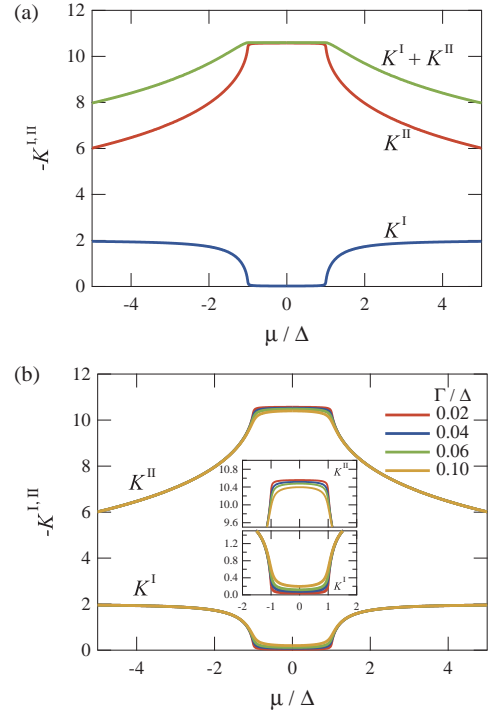}
\caption{(Color online) (a) Chemical potential dependence of the SHC ($K^{\rm I, II}$) for $\Gamma/\D=0.01$ and $E_{\rm c}/\D=100$ at zero temperature.
(b) $K^{\rm I, II}$ for different damping rates: $\Gamma / \D = 0.02, 0.04, 0.06, 0.10$. The inset is the enlarged plot.}
\label{SHE_mu}
\end{center}
\end{figure}

	As for the integration with respect to $\bk$ and $\ve$, we carry out it with respect to $\bk$ first, and then to $\ve$, since the final results are simpler than that obtained in otherwise.\cite{note2}
	Finally, we obtain the results for the spin-Hall conductivity in the form
	\begin{align}
		\sigma_{{\rm s}jk}^i 
	&= \frac{me }{4\pi^2 \hbar^2}\frac{\epsilon_{kji} \left(\alpha_{jk}^2 - \alpha_{jj}\alpha_{kk}\right)}{\sqrt{\det \hat{\alpha}/\D}}
	 \left(K^{\rm I}  + K^{\rm II}  \right),
	 \label{eq_shc}
	\\
	K^{\rm I} &=\int_{-\infty}^{\infty}\!\! d\ve \frac{d f(\ve)}{d\ve}
	\left[ \frac{\sqrt{\ve_+^2 -\D^2}}{\ve}-\frac{\sqrt{\ve_-^2 -\D^2}}{\ve} \right],
	\label{K1}\\
	K^{\rm II} &=\int_{-\infty}^{\infty}\!\! d\ve f(\ve)
	\left[ \frac{1}{\sqrt{\ve_+^2 - \D^2}}-\frac{1}{\sqrt{\ve_-^2 - \D^2}} \right].
	\label{K2}
\end{align}
	(The branch cut of the square root is taken along the positive real axis.)
	The $K^{\rm I}$-term denotes the contribution from $C_2$ and $C_3$, and $K^{\rm II}$-term from $C_1$ and $C_4$, both of which are completely the same as that was obtained for the isotropic Wolff Hamiltonian.\cite{Fuseya2012b}
	Therefore, $\sigma_{{\rm s}jk}^i$ is essentially the same as that of the isotropic model (as is shown in Fig. \ref{SHE_mu}) except for the prefactor.
	Both $K^{\rm I}$ and $K^{\rm II}$ hardly depend on $\Gamma$, reflecting that they are interband contributions.\cite{Fuseya2012b}
	For the clean limit, $\Gamma / \D \to 0$,  we have
	\begin{align}
		K^{\rm I} &= \left\{
		\begin{array}{cc}
		\displaystyle	-\frac{2\sqrt{\mu^2 - \D^2}}{|\mu|} & (|\mu|>\D)
		\\
		\\
		0 & (|\mu|<0)
		\end{array}
		\right. ,
		\label{K1clean}
		\\
		K^{\rm II}&=\left\{
		\begin{array}{cc}
		\displaystyle	-2\ln \left( \frac{2E_{\rm c}}{|\mu| + \sqrt{\mu^2 - \D^2}} \right) & (|\mu|>\D)
		\\
		\\
		\displaystyle 	-2\ln \left( \frac{2E_{\rm c}}{\D}\right) & (|\mu|<\D)
		\end{array}
		\right. ,
		\label{K2clean}
\end{align}
	where $E_{\rm c}$ is the cutoff energy for the $\ve$-integration, and we discarded $\mathcal{O}(\D^2 / E_{\rm c}^2)$ term.

	The spin-Hall conductivity takes its maximum value in the insulating region, i.e., the spin-Hall insulator is achieved.
	The maximum value logarithmically diverges as $\D/E_{\rm c}\to0$.
	It is expected that the spin-current can flow without any Joule heating in the insulating region, $|\mu|<\D$.\cite{note3}

	Note that the present spin-current is in the bulk and not along the surface.
	The magnitude of spin-Hall conductivity depends on $\D/E_{\rm c}$ and does not take a universal value as in SHE in the two-dimensional electron system with Rashba spin-orbit coupling\cite{Sinova2004}, nor quantized as in the quantum SHE.\cite{Kane2005a}

\section{Orbital susceptibility}\label{Orbital susceptibility}

	The formula for the orbital susceptibility, which is exact for Bloch electrons in general, is given by\cite{Fukuyama1971}
\begin{align}
	\chi ^z= \frac{\hbar^2 e^2}{2 c^2}k_{\rm B}T \sum_{n \bk} {\rm Tr}
	\left[ \scr{G} v_y \scr{G} v_x \scr{G} v_y \scr{G} v_x \right],
	\label{chi}
\end{align}
	for a magnetic field $B\parallel z$.
	(Note that, in eq. (\ref{chi}), the spin multiplicity 2 is included by the trace, in contrast to eq. (2.15a) of Ref. \citen{Fukuyama1971}.)
	The matrix elements of $\scr{G} v_y \scr{G} v_x \scr{G} v_y \scr{G}v_x $ becomes
	\begin{align}
	\scr{G} v_y \scr{G} v_x \scr{G} v_y \scr{G} v_x
	&=\frac{M(\bk, \Im \tve_n) }{\left\{ (\Im \tve_n)^2 - E_k^2\right\}^4}, 
	\end{align}
	and 
	\begin{align}
	M(\bk, \Im \tve_n)
	&=
	2\left[ 
	\left\{\hbar^2 (\bk \cdot \bgam) \gamma_y (\bk \cdot \bgam) \gamma_x 
		 +\left\{ (\Im \tilde{\ve}_n)^2 -\D^2 \right\} \gamma_y \gamma_x \right\}^2
	\right]
	\nonumber\\
	&+
	2\hbar^2 \left\{ (\Im \tilde{\ve}_n)^2-\D^2 \right\}
	{\rm Tr}\left[
	\left\{
	(\bk \cdot \bgam) \gamma_y \gamma_x + \gamma_y (\bk \cdot \bgam) \gamma_x
	\right\}^2
	\right]
	\nonumber\\
	&=2\Bigl[ F_{\rm I} + 
	\{ (\Im \ve_n)^2 -\D^2\} \left(F_{\rm II} + F_{\rm III} + F_{\rm IV} + F_{\rm V} \right)
	\nonumber\\
	& + \{ (\Im \ve_n)^2 -\D^2\}^2 F_{\rm VI} \Bigr],
	\\
	F_{\rm I}&= \hbar^4 (\bk \cdot \bgam) \gamma_y (\bk \cdot \bgam) \gamma_x (\bk \cdot \bgam) \gamma_y (\bk \cdot \bgam) \gamma_x ,
	\\
	F_{\rm II}&=
	2\hbar^2 (\bk \cdot \bgam) \gamma_y (\bk \cdot \bgam) \gamma_x \gamma_y \gamma_x ,
	\\
	F_{\rm III}&=
	2\hbar^2 (\bk \cdot \bgam) \gamma_x (\bk \cdot \bgam) \gamma_y \gamma_x \gamma_y ,
	\\
	F_{\rm IV}&=
	\hbar^2 (\bk \cdot \bgam) \gamma_y \gamma_x (\bk \cdot \bgam) \gamma_y \gamma _x ,
	\\
	F_{\rm V}&=
	\hbar^2 (\bk \cdot \bgam) \gamma_x \gamma_y (\bk \cdot \bgam) \gamma_x \gamma _y ,
	\\
	F_{\rm VI}&= \gamma_y \gamma_x \gamma_y \gamma_x .
	\end{align}
	The details of the calculation are given in Appendix\ref{App_Matrix}.
	%
	%
	
	Again, we rotate the $\bk$-space along the spin-space by $\tilde{k}_\mu = \bk \cdot \bm{W}(\mu)$.
	The following relations can be derived by this transformation:
	\begin{align}
	\bk \cdot \bgam &= \sum_\mu \tilde{k}_\mu \sigma_\mu,
\end{align}
	and the commutation relations
	\begin{align}
		(\bk \cdot \bgam) \gamma_i + \gamma_i (\bk \cdot \bgam) &=2\sum_\mu \tilde{k}_\mu W_i (\mu) \equiv 2R_i ,
		\label{eq52}
		\\
		\gamma_i \gamma_j + \gamma_j \gamma_i &= 2\D \alpha_{ij}.
\end{align}
	By using these commutation relations, we can carry out the traces of $F_{\rm I} \sim F_{\rm VI}$ as follows:	
	\begin{align}
	{\rm Tr} \left[ F_{\rm I} \right] 
		&=2\hbar^4 \left[
		8 R_x^2 R_y^2 -8\D \alpha_{xy} \tilde{k}^2 R_x R_y + \D^2 \tilde{k}^4 (2\alpha_{xy}^2 -\alpha_{xx}\alpha_{yy})
		\right],
	\\
	{\rm Tr} \left[ F_{\rm II} \right] 
		&=2\hbar^2 \left[
	8\D \alpha_{xy} R_x R_y - 4\D \alpha_{xx} R_y^2 - 2\D^2 \tilde{k}^2 (2\alpha_{xy}^2 - \alpha_{xx}\alpha_{yy})
	\right],
	\\
	{\rm Tr} \left[ F_{\rm III} \right] 
		&=2\hbar^2 \left[
		8\D \alpha_{xy} R_x R_y - 4\D \alpha_{yy} R_x^2 - 2\D^2 \tilde{k}^2 (2\alpha_{xy}^2 - \alpha_{xx}\alpha_{yy})
		\right],
	\\
	{\rm Tr} \left[ F_{\rm IV} \right] 
		&=4\hbar^2 \D \alpha_{yy}R_x^2 - \frac{1}{2}{\rm Tr} \left[F_{\rm II} \right],
	\\
	{\rm Tr} \left[ F_{\rm V} \right] 
		&=4\hbar^2 \D \alpha_{xx}R_y^2 - \frac{1}{2}{\rm Tr}\left[F_{\rm III}\right],
	\\
	{\rm Tr} \left[ F_{\rm VI} \right] 
		&=2\D^2 (2 \alpha_{xy}^2 - \alpha_{xx}\alpha_{yy}).
\end{align}
	Consequently, the trace of the matrix elements becomes
\begin{align}
	&{\rm Tr} \left[ M(\bk, \Im \tve_n) \right]
	\nonumber\\
	&=
	-4\D^2 \alpha_{xx}\alpha_{yy} \left\{(\Im \tve_n)^2 -E_{\tilde{k}}^2 \right\}^2 + 8\left[ \D \alpha_{xy} \left\{(\Im \tve_n)^2 -E_{\tilde{k}}^2 \right\}+ 2\hbar^2 R_x R_y \right]^2
	\label{eq63}
\end{align}
	where we used the relation $\D^2 = E_k^2 - \hbar^2 \tilde{k}^2$.
	The contributions from $\bm{R}$, eq. (\ref{eq52}), are
	\begin{align}
		R_x R_y &= \frac{4\pi}{3}\D \tilde{k}^2 \alpha_{xy},
		\\
		R_x^2 R_y^2 &= \frac{4\pi}{15}\D^2 \tilde{k}^4 (2\alpha_{xy}^2 + \alpha_{xx} \alpha_{yy} ),
\end{align}
	where the terms linear in $\tilde{k}_{1, 2, 3}$ are discarded since they vanish in the integration with respect to $\tilde{k}_{1,2,3}$.
	After the integration with respect to the solid angle, we obtain
	\begin{align}
		{\rm Tr} \left[ M(\bk, \epsilon) \right]
		&=16\pi \D^2 \Biggl[
		(2\alpha_{xy}^2 -\alpha_{xx}\alpha_{yy} )\left\{(\Im \ve_n)^2 -E_k^2 \right\}^2
		\nonumber\\ &
		+\frac{8}{3}\alpha_{xy}^2 \hbar^2 \tilde{k}^2 \left\{(\Im \ve_n)^2 -E_k^2 \right\}
		+ \frac{8}{15}(2\alpha_{xy}^2 + \alpha_{xx}\alpha_{yy}) \hbar^4 \tilde{k}^4
		\Biggr].
\end{align}
	The integrations associated with the Green's function are as follows:
	\begin{align}
		k_{\rm B}T&\sum_n \sum_{\tilde{k}_{1,2,3}} \frac{1}{\left[ (\Im \tve_n )^2 -E_{\tilde{k}}^2 \right]^2}
		\nonumber \\
		&=-\frac{1}{16\pi^2 \hbar^3} \int_{-\infty}^{\infty}\!\! d \ve \, f(\ve)
		\left( \frac{1}{\sqrt{\ve_+^2 - \D^2}} - \frac{1}{\sqrt{\ve_-^2 - \D^2}} \right),
		\\
		k_{\rm B}T&\sum_n \sum_{\tilde{k}_{1,2,3}} \frac{\hbar^2 \tilde{k}^2}{\left[ (\Im \tve_n )^2 -E_{\tilde{k}}^2 \right]^3}
		\nonumber \\
		&=\frac{3}{64\pi^2 \hbar^3} \int_{-\infty}^{\infty}\!\! d \ve \, f(\ve)
		\left( \frac{1}{\sqrt{\ve_+^2 - \D^2}} - \frac{1}{\sqrt{\ve_-^2 - \D^2}} \right),
		\\
		k_{\rm B}T&\sum_n \sum_{\tilde{k}_{1,2,3}} \frac{\hbar^4 \tilde{k}^4}{\left[ (\Im \tve_n )^2 -E_{\tilde{k}}^2 \right]^4}
		\nonumber \\
		&=-\frac{5}{128\pi^2 \hbar^3} \int_{-\infty}^{\infty}\!\! d \ve \, f(\ve)
		\left( \frac{1}{\sqrt{\ve_+^2 - \D^2}} - \frac{1}{\sqrt{\ve_-^2 - \D^2}} \right).
\end{align}
	(The contours are depicted in Fig. \ref{contours} (b).)
	Combining the trace of the matrix elements and integration with respect to $\bk$, we have
	\begin{align}
		k_{\rm B}T\!\! &\sum_{n, \tilde{k}_{1,2,3}} {\rm Tr}\left[
		\scr{G} v_y \scr{G} v_x \scr{G} v_y \scr{G} v_x
		\right]
		=k_{\rm B}T \!\!\sum_{n, \tilde{k}_{1,2,3}} \frac{ {\rm Tr} [M( \bk, \Im \tve_n )]}{\{ (\Im \tve_n)^2 - E^2 \}^4}
		\nonumber\\
		&=\frac{\D^2}{4\pi^2 \hbar^3} \Biggl[
		-(2\alpha_{xy}^2 -\alpha_{xx}\alpha_{yy} )
		+2\alpha_{xy}^2
		- \frac{1}{3}(2\alpha_{xy}^2 + \alpha_{xx}\alpha_{yy}) 
		\Biggr] K^{\rm II}
		\nonumber\\
		&=
		-\frac{\D^2}{6\pi^2\hbar^3}\left( \alpha_{xy}^2 - \alpha_{xx}\alpha_{yy} \right) K^{\rm II}.
\end{align}
	The functional form of $K^{\rm II}$ is the same as that for the spin-Hall conductivity, eq. (\ref{K2}), and so the same as that for the isotropic Wolff Hamiltonian.

	Finally, we obtain the orbital susceptibility for $B\parallel i$ in the form
	\begin{align}
		\chi^{i} = -\frac{e^2 }{12\pi^2\hbar c^2 }\frac{\left( \alpha_{jk}^2 - \alpha_{jj}\alpha_{kk} \right)}{\sqrt{\det  \hat{\alpha}/\D}} K^{\rm II}.
		\label{eq_chi}
\end{align}
	($j$ and $k$ are perpendicular to $i$, e.g., $j=y$ and $k=x$ for $i=z$.)
	Now we have an exact relationship between $\sigma_{{\rm s}jk}^{i}$ and $\chi^{i}$ for the anisotropic Wolff Hamiltonian in the insulating state ($|\mu |<\D $):
	\begin{align}
		\sigma_{{\rm s}jk}^{i}= \frac{3mc^2}{\hbar e}\epsilon_{ijk} \chi^{i}.
		\label{relation}
\end{align}
	Surprisingly, the relation for the isotropic Wolff Hamiltonian, eq. (\ref{eq1}), is valid also for the anisotropic Wolff Hamiltonian.
	The relation eq. (\ref{relation}) holds for various Dirac electron systems whose chemical potential is located in the band gap, since the anisotropic Wolff Hamiltonian is an effective Hamiltonian common to Dirac electron systems as is mentioned.
	There should be some physical reason behind this clear relationship between the spin-Hall conductivity and orbital susceptibility, which unfortunately we have not yet been able to clarify.
	
	The results of the spin-Hall conductivity eq. (\ref{eq_shc}) and the orbital susceptibility eq. (\ref{eq_chi}), and then eq. (\ref{relation}), are valid even for such two-dimensional systems as $\alpha$-ET$_2$I$_3$\cite{Kobayashi2007}, where the energy dispersion is two-dimensional and the spin-space is three dimensional, since the dimensionality of energy dispersion is taken into account in terms of the inverse mass-tensor. (See also Appendix \ref{App_2D}.)

\section{Discussion}\label{Discussion}
\begin{figure}
\begin{center}
\includegraphics[width=40mm]{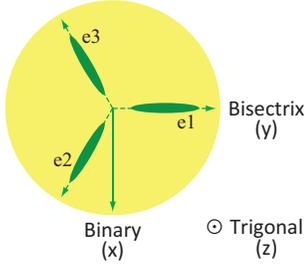}
\end{center}
\caption{(Color online) Schematic illustration of electron Fermi surfaces of Bi.}
\label{3FS}
\end{figure}
	%
	Here we give a quantitative estimation of the orbital susceptibility and the spin-Hall conductivity for Bi and its alloys with Sb.
	The electronic structure of Bi is well known:\cite{Dresselhaus1971,Edelman1976,Issi1979,Smith1964,ZZhu2011,ZZhu2012} 
	There are three electron ellipsoids at $L$-points and one hole ellipsoid at $T$-point.
	The dominant contribution to the transport coefficients is that from electrons at $L$-points, since the mass of electrons are much smaller than that of holes.
	The electron ellipsoids are highly elongated  with ratio of major to minor axes $\sim 15:1$.
	Each electron ellipsoid can be transformed into one another with the rotation by $120^\circ$ about the trigonal axis as is shown in Fig. \ref{3FS}.
	We label the electron ellipsoid along the bisectrix axes as ``e1", and the other as ``e2" and ``e3" (Fig. \ref{3FS}).
	The inverse mass-tensor of e1 is given by\cite{Dresselhaus1971,Issi1979,Smith1964,ZZhu2011}
	\begin{align}
		%
		\hat{\alpha}_{\rm e1}=
		\begin{pmatrix}
			\alpha_{1} & 0 & 0 \\
			0 & \alpha_{2} & \alpha_{4} \\
			0 & \alpha_{4} & \alpha_{3}
		\end{pmatrix}.
\end{align}
	%
	%
	The inverse mass-tensors of e2 and e3 are then obtained by the $120^\circ$ rotation as
	\begin{align}
		\hat{\alpha}_{\rm e2, e3} = \frac{1}{4}
		\begin{pmatrix}
			\alpha_1 + 3 \alpha_2 & \pm \sqrt{3}(\alpha_1 - \alpha_2) & \pm 2\sqrt{3} \alpha_4 \\
			 \pm \sqrt{3}(\alpha_1 - \alpha_2)  & 3\alpha_1 + \alpha_2 & -2 \alpha_4 \\
			 \pm 2\sqrt{3} \alpha_4 & -2 \alpha_4 & 4 \alpha_3
		\end{pmatrix}.
\end{align}
	The Gaussian curvatures for each ellipsoid are obtained as 
	\begin{align}
		\alpha_{xy}^2 - \alpha_{xx}\alpha_{yy} &= 
		\left\{
		\begin{array}{@{\,}ll}
			-\alpha_1 \alpha_2 & \mbox{(e1)} \\
			-\alpha_1 \alpha_2 & \mbox{(e2, e3)} 
		\end{array}
		\right.,
		\\
		\alpha_{yz}^2 - \alpha_{yy}\alpha_{zz} &= 
		\left\{
		\begin{array}{@{\,}ll}
			\alpha_4^2 -\alpha_2 \alpha_3 & \mbox{(e1)} \\
			\frac{1}{4}\alpha_4^2 - \frac{1}{4}(3\alpha_1 + \alpha_2)\alpha_3 & \mbox{(e2, e3)} 
		\end{array}
		\right.,
		\\
		\alpha_{zx}^2 - \alpha_{zz}\alpha_{xx} &= 
		\left\{
		\begin{array}{@{\,}ll}
			-\alpha_3 \alpha_1 & \mbox{(e1)} \\
			\frac{3}{4}\alpha_4^2 - \frac{1}{4}(\alpha_1 + 3\alpha_2)\alpha_3 & \mbox{(e2, e3)} 
		\end{array}
		\right. .
\end{align}
	Then the total contribution from electrons is given by the summation of the each Gaussian curvature:
	\begin{align}
		\left(\alpha_{xy}^2 - \alpha_{xx}\alpha_{yy}\right)_{\rm total} &= -3 \alpha_1 \alpha_2 \equiv \kappa_{\parallel},\\
		\left(\alpha_{yz}^2 - \alpha_{yy}\alpha_{zz}\right)_{\rm total} &= \frac{3}{2}\alpha_4^2 -\frac{3}{2}(\alpha_1 + \alpha_2) \alpha_3 \equiv \kappa_{\perp},\\
		\left(\alpha_{zx}^2 - \alpha_{zz}\alpha_{xx}\right)_{\rm total} &= \frac{3}{2}\alpha_4^2 -\frac{3}{2}(\alpha_1 + \alpha_2) \alpha_3 =\kappa_{\perp}.
\end{align}
	The Gaussian curvature for the binary plane, $(\alpha_{yz}^2 - \alpha_{yy}\alpha_{zz})_{\rm total}$, is the same as that for the bisectrix plane, $(\alpha_{zx}^2 - \alpha_{zz}\alpha_{xx})_{\rm total}$.

	The components of the mass-tensor, $\hat{m}$, obtained by the recent angle resolved Landau level measurements\cite{ZZhu2011,ZZhu2012} and the corresponding components of $\hat{\alpha}$ ($=\hat{m}^{-1}$) are listed in Table \ref{t1}.
	%
\begin{table}
\caption{Parameters for the mass tensor $\hat{m}$\cite{ZZhu2011} (in the unit of the bare electron mass $m$) and the inverse mass tensor $\hat{\alpha}$ (in the unit of $m^{-1}$).}
\label{t1}
\begin{center}
\begin{tabular}{lcccc}
\hline
\hline
$i$ & $1$ & $2$ & $3$ & $4$ \\
\hline
mass ($m_i$) & 0.00124 & 0.257 & 0.00585 & -0.0277 \\
inverse mass ($\alpha_i$) & 806 & 7.95 & 349 & 37.6 \\
\hline
\end{tabular}
\end{center}
\end{table}
	%
	By using these values, we can estimate the total Gaussian curvature as follows:
	\begin{align}
		\kappa_{\parallel} m^2&= -1.92 \times 10^4 ,\\
		\kappa_{\perp} m^2 &= -4.24 \times 10^5.
\end{align}
	%
\begin{figure}
\begin{center}
\includegraphics[width=80mm]{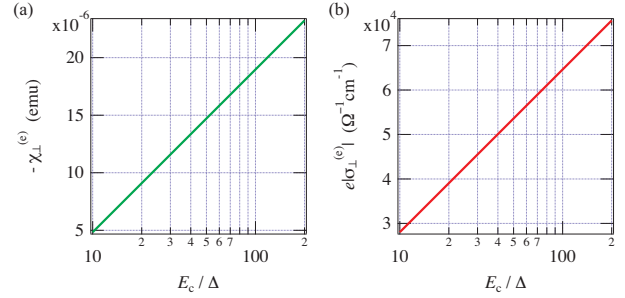}
\end{center}
\caption{(Color online) Energy cutoff $E_{\rm c}$ dependence of (a) orbital susceptibility and (b) spin-Hall conductivitiy.}
\label{Ec}
\end{figure}
	%
	Now we can estimate the orbital susceptibility (electron contribution only) at zero temperature in the clean limit ($\Gamma /\D \to 0$) as
	\begin{align}
		\chi_{\perp}^{\rm (e)}&=
		-\frac{e^2 \kappa_\perp}{12\pi^2 \hbar c^2} 
		\sqrt{\frac{\D_{\rm e}}{\det \hat{\alpha}}}
		K^{\rm II}(\mu_{\rm e})
		\nonumber\\
		&=
		\left( 9.27-6.11\times \ln \frac{E_{\rm c}}{\D_{\rm e}} \right) \times 10^{-6} \,\, \mbox{emu},
\end{align}
	where the band gap at $L$-point is $\D_{\rm e}=7.65$ meV, and the chemical potential of electron is $\mu_{\rm e}=35.3$ meV for pure Bi\cite{Smith1964,ZZhu2011}.
	Only parameter is $E_{\rm c}$.
	However, the order of magnitude of $\chi_{\perp}^{\rm (e)} (\sim 10^{-5}$emu) is unchanged even if we change $E_{\rm c}$ largely as is shown in Fig. \ref{Ec} (a).
	If we chose $E_{\rm c}/\D_{\rm e}=100$, we have $\chi_{\perp}= -1.89 \times 10^{-5}$emu, which agrees well with the experimental results $\chi_\perp=-1.94 \times 10^{-5}$emu obtained by Otake {\it et al.}\cite{Otake1980,note1}.
	%
	%
	When we carry out the same estimation for $\chi_{\parallel}^{\rm (e)}$, we have $\chi_{\parallel}^{\rm (e)}=-8.56 \times 10^{-7}$emu, which is smaller than the experimental result $\chi_\parallel=-1.41 \times 10^{-5}$emu.
	This discrepancy is due to the large contribution from holes in this direction.\cite{Fukuyama1970}
	Unfortunately, it is rather difficult to give an accurate estimation of the contribution from hole, since the magnitude of the energy gap is unclear at $T$-point.

	The spin-Hall conductivity can be estimated at zero temperature in the $\Gamma/\D \to 0$ limit as
	\begin{align}
		e\sigma_{{\rm s}\perp}^{\rm (e)}&=\frac{me^2 \kappa_\perp}{4\pi^2 \hbar^2}\sqrt{\frac{\D_{\rm e}}{\det \hat{\alpha}}}
		\left[K^{\rm I}(\mu_{\rm e}) + K^{\rm II}(\mu_{\rm e}) \right]
		\nonumber\\
		&=\left(-0.855+1.58\times \ln \frac{E_{\rm c}}{\D_{\rm e}}\right)\times 10^4 \,\, \Omega^{-1}{\rm cm}^{-1} .
		\label{eq85}
\end{align}
	Again, the order of $e|\sigma_{{\rm s}\perp}^{\rm (e)}| (\sim 10^4 \Omega^{-1}{\rm cm}^{-1})$ is unchanged even if we change $E_{\rm c}$ in a wide range (Fig. \ref{Ec} (b)).
	If we chose the same value as in $\chi_\perp^{\rm (e)}$, $E_{\rm c}/\D_{\rm e}=100$, we have $e|\sigma_{{\rm s}\perp}^{\rm (e)}|=6.44 \times 10^4\, \Omega^{-1}{\rm cm}^{-1}$. 

	We can also estimate the magnitude of $\sigma_{{\rm s}xy}$ directly from the experimental value of the orbital susceptibility by the relationship (eq. (\ref{relation})):
	\begin{align}
		e|\sigma_{{\rm s}xy}| =(2.59 \times 10^9 ) \chi \,\, \Omega^{-1}{\rm cm}^{-1},
		\label{eq86}
\end{align}
	where $3mc^2/\hbar = 2.59 \times 10^9 \,\, \Omega^{-1}{\rm cm}^{-1}$.
	To be precise, the contribution from $K^{\rm I}$ is neglected here, so that this is exact only in the insulating case, and the actual value of $e|\sigma_{{\rm s}xy}|$ should be larger for the metallic case.
	However, $K^{\rm I}$ gives almost constant contribution with respect to $\mu$, and it is smaller than $K^{\rm II}$, so that the general situation will not vary even if we take into account the contribution from $K^{\rm I}$.
	Therefore, eq. (\ref{eq86}) should be useful to estimate the magnitude of $e|\sigma_{{\rm s}xy}|$ from the experimental value of $\chi$ even for the (semi-) metallic case, such as pure Bi.
	From the magnetic susceptibility of pure Bi at room temperature\cite{Otake1980}, $\chi_{\perp}^{\rm (rt)}=1.43 \times 10^{-5}$ emu, we obtain $e|\sigma_{{\rm s}\perp}^{\rm (rt)}| = 3.70 \times 10^4 \Omega^{-1} {\rm cm}^{-1}$ at room temperature, which is consistent with the result of eq. (\ref{eq85}).
	This is quite large compared to the conventional spin-Hall conductivity.
	For example, the spin-Hall conductivity of Pt, the typical material which exhibits the SHE, is as $e|\sigma_{{\rm s}xy}|\simeq 2.4 \times 10^2 \Omega^{-1} {\rm cm}^{-1}$ at room temperature\cite{Kimura2007,Guo2008,Kontani2008,Tanaka2008}.
	This is about $10^4$ times larger than the value reported in $n$-type semiconductors.
	%
	%
	%
	Furthermore, the spin-Hall conductivity would be enhanced by alloying Bi with Sb\cite{Fuseya2012b}, since the increase of the diamagnetism has been already measured.\cite{Wehrli1968}

\section{Summary}\label{Summary}

	We have derived an important relationship between the spin-Hall conductivity $\sigma_{{\rm s}jk}^i$ and orbital susceptibility $\chi^i$ for the anisotropic Wolff Hamiltonian, which is an effective Hamiltonian common to Dirac electrons in solids.
	For both $\sigma_{{\rm s}jk}^i$ and $\chi^i$, the anisotropy appears only in their prefactors, which is given by the Gaussian curvature $(\alpha_{jk}^2 - \alpha_{jj}\alpha_{kk})$, and then the relationship between $\sigma_{{\rm s}jk}^i$ and $\chi^i$ discovered for the isotropic case in the insulating state, $\sigma_{{\rm s}jk}^i = (3mc^2/\hbar e) \epsilon_{ijk} \chi^i$ (eq. (\ref{relation})), is also valid even for the anisotropic case. 
	Note that the relationship (\ref{relation}) is exact only for the insulating case, but is useful for the estimation even for the (semi-) metallic case, such as pure bismuth.
	This is because the contribution that appears in the metallic state, i.e., $K^{\rm I}$-term (eq. (\ref{K1})), is smaller than the dominant term of $K^{\rm II}$ (eq. (\ref{K2})), and it does not change the general situation.
	%
	%
	
	We have evaluated the magnitude of $\chi^i$ and $\sigma_{{\rm s}jk}^i$ for pure bismuth by using the mass-tensor values obtained by the recent angle resolved Landau level measurements.\cite{ZZhu2011,ZZhu2012}
	We obtained $\chi_{\perp}=-1.89\times 10^{-5}$emu at zero temperature for $E_{\rm c}/\D_{\rm e} = 100$, which gives good agreement with the experimental results.\cite{Otake1980}
	Based on this, we obtained $e |\sigma_{{\rm s}\perp}|\simeq 4\times 10^4 \Omega^{-1}{\rm cm}^{-1}$ at room temperature, which is about $10^2$ times larger than that of Pt.\cite{Kimura2007,Guo2008,Kontani2008,Tanaka2008}
	The magnitude of $e \sigma_{{\rm s}\perp}$ can be increased by lowering temperature or by the substitution of Bi atoms by Sb atoms.

\begin{acknowledgment}

The authors would like to thank K. Behnia and H. Harima for fruitful discussions and comments.
This  work is supported by Grants-in-Aid for Scientific Research on ``Dirac electrons in Solids" (No. 24244053).
Y. F. is also supported by Young Scientests B (No. 25870231).

\end{acknowledgment}

\appendix
\section{Gaussian curvature}\label{Gaussian}

	The spin-Hall conductivity has a coefficient
	\begin{align}
		\sum_{\lambda \mu \nu} \epsilon_{\lambda \mu \nu} Q_i (\lambda) W_j (\mu) W_k (\nu),
		\label{QWW}
\end{align}
	as eq. (\ref{eq39}), where $\bm{Q}(\lambda )=\Omega^{-1}\bm{W}(\mu)\times \bm{W}(\nu)$.
	(The values $(\lambda, \mu, \nu)$ occur in the cyclic order of $(1, 2, 3)$.)
	This factor is written in terms of $\bm{W}(\mu)$, which does not corresponds directly to the physical quantities obtained by experiments.
	In order to make our theory more practical one, we need to rearrange eq. (\ref{QWW}) in terms of the quantities directly corresponds to experiments.
	In a following manner, we can express eq. (\ref{QWW}) in terms of the inverse mass-tensor, which can be determined experimentally.

	First, we examine the case of $(i, j, k)=(z, y, x)$, namely, the electric field is along $x$-axis, and the spin current is along $y$-axis, where the spin-magnetic moment of the spin current is along $z$-axis.
	In this case, we have
	\begin{align}
		\sum_{\lambda \mu \nu} &\epsilon_{\lambda \mu \nu} Q_z (\lambda) W_y (\mu) W_x (\nu)
		\nonumber\\
		&=
		\Omega^{-1}\Bigl[
		\left\{ W_x (2) W_y (3) - W_y(2) W_x (3)\right\} \left\{ W_y (2) W_x (3) - W_y (3) W_x (2) \right\}
		\nonumber\\
		&+
		\left\{ W_x (3) W_y (1) - W_y(3) W_x (1)\right\} \left\{ W_y (3) W_x (1) - W_y (1) W_x (3) \right\}
		\nonumber\\
		&+
		\left\{ W_x (1) W_y (2) - W_y(1) W_x (2)\right\} \left\{ W_y (1) W_x (2) - W_y (2) W_x (1) \right\}
		\Bigr]
		\nonumber\\
		&=
		\Omega^{-1}
		\Bigl[
		\left\{ W_x (1) W_y (1) + W_x (2) W_y (2) + W_x (3) W_y (3) \right\}^2
		\nonumber\\
		&-
		\left\{ W_x^2 (1) + W_x^2 (2) + W_x^2 (3) \right\}
		\left\{ W_y^2 (1) + W_y^2 (2) + W_y^2 (3) \right\}
		\Bigr]
		\nonumber\\
		&=\frac{\D^2}{\Omega}\left( \alpha_{xy}^2 - \alpha_{xx}\alpha_{yy} \right).
	\end{align}
	For the other direction, the similar derivation yields:
	\begin{align}
		\sum_{\lambda \mu \nu} \epsilon_{\lambda \mu \nu}
		Q_i (\lambda) W_j (\mu) W_k (\nu)
		= \frac{\D^2}{\Omega}\epsilon_{kji}
		\left( \alpha_{jk}^2 - \alpha_{jj} \alpha_{kk} \right),
\end{align}
	for $i, j, k$ are perpendicular with each other.

	Generally speaking, the factor $\sum_{\lambda \mu \nu} Q_i (\lambda) W_j (\mu) W_k (\nu)$ can be finite for the case that two of the three directions are the same.
	(Note that it is obviously zero for $i=j=k$.)
	Actually, we can show
	\begin{align}
		\kappa_{iji}\equiv \sum_{\lambda \mu \nu} \epsilon_{\lambda \mu \nu} Q_i (\lambda) W_j (\mu) W_i (\nu)
		=\frac{\D^2}{\Omega}\epsilon_{ijk}\left( \alpha_{jj}\alpha_{ki} - \alpha_{jk} \alpha_{ji} \right),
		\label{QWWiji}
\end{align}
	where $k$ is perpendicular to both $i$ and $j$.
	However, this is expected to vanish in the real materials with the inversion symmetry.
	For example, when the axes of the equal energy surface is along the $x, y, z$-axes, $\alpha_{ij}=0$, so that $\kappa_{iji}=0$.
	Even in the case of the equal energy surface is tilted from the $x, y, z$-axes, the total coefficient vanishes for the spacially symmetric crystals as follows.
	Here we consider the three electron ellipsoids of bismuth discussed in \S\ref{Discussion} of the main body of the present paper.
	The followings are the possible combinations:
	\begin{align}
		\kappa_{zxz}=\alpha_{xx}\alpha_{yz}-\alpha_{xy}\alpha_{xz} &=
		\left\{
		\begin{array}{@{\,}ll}
			\alpha_1 \alpha_4 & {\rm (e1)}\\
			-\frac{1}{2}\alpha_1 \alpha_4 & {\rm (e2, e3)}
		\end{array}
		\right. ,
		\\
		\kappa_{xyx}=\alpha_{yy}\alpha_{zx}-\alpha_{yz}\alpha_{yx} &=
		\left\{
		\begin{array}{@{\,}ll}
			0 & {\rm (e1)}\\
			\pm \frac{\sqrt{3}}{2}\alpha_1 \alpha_4 & {\rm (e2, e3)}
		\end{array}
		\right. ,
		\\
		\kappa_{yzy}=\alpha_{zz}\alpha_{xy}-\alpha_{zx}\alpha_{zy} &=
		\left\{
		\begin{array}{@{\,}ll}
			0 & {\rm (e1)}\\
			\pm \frac{\sqrt{3}}{4}(\alpha_1 \alpha_3 - \alpha_2 \alpha_3 + \alpha_4^2) & {\rm (e2, e3)}
		\end{array}
		\right. .
\end{align}
	For each case, the total $\kappa_{iji}$ is zero.
	Other combinations can be re-expressed as the above three types, e.g. $\kappa_{xxy}=-\kappa_{xyx}$ and $\kappa_{xxz}=-\kappa_{xzx}=\kappa_{yzy}$.
	Consequently, the SHE occurs only when the directions of the electric field, the spin current, and the spin magnetic-moment are perpendicular to each other.
	 So far, such a configuration has been studied rather phenomenological picture.\cite{Hirsch1999,Murakami2003}
	 Our derivation gives microscopic proof of the configuration for SHE.

\section{Analytic Continuation}\label{App_Analytic}
	Here we describe the details of the analytic continuation for eq. (\ref{eq34}).
	The analytic continuation for the orbital susceptibility can be carried out in the same manner.
	The frequency summation of a function $\mathcal{F} (\Im \ve_n)$ continues analytically to $\mathcal{F}(z)$ as
	\begin{align}
		k_{\rm B}T\sum_n \mathcal{F}(\Im \ve_n) = -\frac{1}{2\pi \Im} \int_C \!\! dz\, f(z) \mathcal{F}(z),
\end{align}
	where $f(\ve)$ is the Fermi distribution function, and $\mathcal{F}(z)$ is an analytic function in the region enclosed by the contour $C$.
	%
	%
	For the transport coefficients, the contour is composed of four contours $C_1 \sim C_4$ displayed in Fig. \ref{contours} (a).
	(For the orbital susceptibility, the contour is composed of two contours $C_{\rm a}$ and $C_{\rm b}$ displayed in Fig. \ref{contours} (b).)
	After the analytic continuation with respect to $\ve_n$ and then to $\omega_\lambda$, each contribution becomes as follows:
	\begin{align}
		 -ek_{\rm B} T &\sum_{n} 
	\frac{  \Im \tilde{\ve}_n - \Im \tilde{\ve}_{n-}}{\{ (\Im \tilde{\ve}_n)^2 -E_{\tilde{k}}^2 \}\{ (\Im \tilde{\ve}_{n-})^2 -E_{\tilde{k}}^2 \}}
	\nonumber\\
	&=
	-\frac{1}{2\pi \Im }\int_{-\infty}^{\infty} \!\! d\ve\, f(\ve)
	\nonumber\\
	&\times
	\Biggl[
	\frac{\hbar \omega}{\left\{ (\ve + \hbar \omega+\Im \Gamma)^2 -E_{\tilde{k}}^2 \right\} 
	\left\{ (\ve+\Im \Gamma )^2 -E_{\tilde{k}}^2 \right\}}
	\tag{$C_1$}
	\nonumber\\
	&-\frac{\hbar \omega+2\Im \Gamma}{\left\{ (\ve + \hbar \omega+\Im \Gamma)^2 -E_{\tilde{k}}^2 \right\} 
	\left\{ (\ve-\Im \Gamma )^2 -E_{\tilde{k}}^2 \right\}}
	\tag{$C_2$}
	\nonumber\\
	&+\frac{\hbar \omega+2\Im \Gamma}
	{\left\{ (\ve +\Im \Gamma)^2 -E_{\tilde{k}}^2 \right\} 
	\left\{ (\ve -\hbar \omega-\Im \Gamma )^2 -E_{\tilde{k}}^2 \right\}}
	\tag{$C_3$}
	\nonumber\\
	&-\frac{\hbar \omega}
	{\left\{ (\ve -\Im \Gamma)^2 -E_{\tilde{k}}^2 \right\} 
	\left\{ (\ve -\hbar \omega -\Im \Gamma )^2 -E_{\tilde{k}}^2 \right\}}
	\tag{$C_4$}
	\Biggr].
\end{align}
	For the dc conductivity, we only need the term linear in $\omega$.
	Then the contribution from $C_1$ and $C_4$ is
	\begin{align}
	-\frac{\hbar \omega}{2\pi \Im }\int_{-\infty}^{\infty} \!\! d\ve\, f(\ve)
	&
	\left[ 
	\frac{1}{\left( \ve_+^2 -E_{\tilde{k}}^2 \right)^2}
	-\frac{1}
	{\left( \ve_-^2 -E_{\tilde{k}}^2 \right)^2}
	\right],
\end{align}
	where $\ve_\pm = \ve \pm \Im \Gamma$, and that from $C_2$ and $C_3$ is
	\begin{align}
		&-\frac{\hbar \omega}{2\pi \Im } \int_{-\infty}^\infty \!\! d\ve \, f(\ve)
		\left\{ -\frac{d}{d\ve}
		\frac{2\Im \Gamma}{\left( \ve_+^2 - E_{\tilde{k}}^2 \right)\left( \ve_-^2 - E_{\tilde{k}}^2 \right)}
		\right\}
	\nonumber\\
	&=-\frac{\hbar \omega}{2\pi \Im } \int_{-\infty}^\infty \!\! d\ve \, 
	\frac{d f(\ve)}{d \ve}
		\frac{2\Im \Gamma}{\left( \ve_+^2 - E_{\tilde{k}}^2 \right)\left( \ve_-^2 - E_{\tilde{k}}^2 \right)},
\end{align}
	where we carried out the integration by parts.

	We first carry out the integration with respect to $\bk$.\cite{note2}
	For $C_1$ and $C_4$, 
	\begin{align}
		&\frac{1}{(2\pi)^3} \int_{-\infty}^{\infty}\!\! d^3 \tilde{k}\,
		\left[
		\frac{1}{\left( \ve_+^2 -E_{\tilde{k}}^2 \right)^2}
	-\frac{1}{\left( \ve_-^2 -E_{\tilde{k}}^2 \right)^2}
	\right]
	\nonumber\\
	&
	=\frac{\Im}{8\pi \hbar^3 \D^{3/2}}
	\left[
	\frac{1}{\sqrt{\ve_+^2 - \D^2}}-\frac{1}{\sqrt{\ve_-^2 - \D^2}}
	\right].
\end{align}
	(The branch cut of the square root is taken along the positive real axis.)
	Here we safely expanded the region of integration from $0 \to \infty$ to $-\infty \to \infty$, and used the residue theorem.
	For $C_2$ and $C_3$, 
	\begin{align}
		&\frac{1}{(2\pi)^3}\int_{-\infty}^{\infty} \!\! d^3 \tilde{k}\,
		\frac{2\Im \Gamma}{\left( \ve_+^2 -E_{\tilde{k}}^2 \right)\left( \ve_-^2 -E_{\tilde{k}}^2 \right)}
		\nonumber\\
		%
		%
		&=\frac{\Im }{8\pi \hbar^3 \D^{3/2}}\frac{\sqrt{\ve_+^2 -\D^2}-\sqrt{\ve_-^2 -\D^2}}{\ve}.
\end{align}

	Finally, we obtain the results for the spin-Hall conductivity in the form
	\begin{align}
		\sigma_{{\rm s}jk}^i 
	&= \frac{me }{4\pi^2 \hbar^2}\frac{\left(\alpha_{jk}^2 - \alpha_{jj}\alpha_{kk}\right)}{\sqrt{\det \hat{\alpha}/\D}}
	 \left(K^{\rm I}  + K^{\rm II}  \right),
	\\
	K^{\rm I} &=\int_{-\infty}^{\infty}\!\! d\ve \frac{d f(\ve)}{d\ve}
	\left[ \frac{\sqrt{\ve_+^2 -\D^2}}{\ve}-\frac{\sqrt{\ve_+^2 -\D^2}}{\ve} \right],\\
	K^{\rm II} &=\int_{-\infty}^{\infty}\!\! d\ve f(\ve)
	\left[ \frac{1}{\sqrt{\ve_+^2 - \D^2}}-\frac{1}{\sqrt{\ve_-^2 - \D^2}} \right].
\end{align}

\section{Matrix Elements for Orbital Susceptibility}\label{App_Matrix}

	Here we note the calculation of the matrix elements for the orbital susceptibility.
	The matrix elements for the orbital susceptibility are the following:
\begin{align}
{\rm Tr}&\left[ \scr{G} v_y \scr{G} v_x \scr{G} v_y \scr{G} v_x\right] \times \left\{ (\Im \tilde{\ve}_n )^2 - E_k^2 \right\}^4
\nonumber\\
&=
2{\rm Tr}\left[ 
\left\{\hbar^2 (\bk \cdot \bgam) \gamma_y (\bk \cdot \bgam) \gamma_x 
	 +\left\{ (\Im \tilde{\ve}_n)^2 -\D^2 \right\} \gamma_y \gamma_x \right\}^2
\right]
\nonumber\\
&+
2\hbar^2 \left\{ (\Im \tilde{\ve}_n)^2-\D^2 \right\}
{\rm Tr}\left[
\left\{
(\bk \cdot \bgam) \gamma_y \gamma_x + \gamma_y (\bk \cdot \bgam) \gamma_x
\right\}^2
\right]
\nonumber\\
&=2{\rm Tr}\biggl[
\hbar^4 (\bk \cdot \bgam) \gamma_y (\bk \cdot \bgam) \gamma_x (\bk \cdot \bgam) \gamma_y (\bk \cdot \bgam) \gamma_x
\nonumber\\
&+\hbar^2 \{(\Im \tilde{\ve}_n )^2 -\D^2\}\Bigl\{
2(\bk \cdot \bgam) \gamma_y (\bk \cdot \bgam) \gamma_x \gamma_y \gamma_x 
+2(\bk \cdot \bgam) \gamma_x (\bk \cdot \bgam) \gamma_y \gamma_x \gamma_y
\nonumber\\
&+
(\bk \cdot \bgam) \gamma_y \gamma_x (\bk \cdot \bgam) \gamma_y \gamma _x
+(\bk \cdot \bgam) \gamma_x \gamma_y (\bk \cdot \bgam) \gamma_x \gamma _y 
\Bigr\}
\nonumber\\&
+\{(\Im \tilde{\ve}_n )^2 -\D^2\}^2 \gamma_y \gamma_x \gamma_y \gamma_x
\biggr]
\end{align}
Here we denote each term as $F_{\rm I} \sim F_{\rm VI}$.

We rotate the $\bk$-space along the spin-space by the transformation
\begin{align}
	\tilde{k}_\mu = \bk \cdot \bW( \mu) =k_x W_x (\mu) + k_y W_y (\mu) + k_z W_z (\mu).
\end{align}
Then we have
\begin{align}
	\bk \cdot \bgam &= k_x \left\{ W_x (1) \sigma_1 + W_x (2) \sigma_2 + W_x (3) \sigma_3 \right\}
	\nonumber\\
	&+k_y \left\{ W_y (1) \sigma_1 + W_y (2) \sigma_2 + W_y (3) \sigma_3 \right\}
	\nonumber\\
	&+k_z \left\{ W_z (1) \sigma_1 + W_z (2) \sigma_2 + W_z (3) \sigma_3 \right\}
	\nonumber\\
	&= \sum_\mu \tilde{k}_\mu \sigma_\mu ,
\end{align}
	and
	\begin{align}
		(\bk \cdot \bgam) (\bk \cdot \bgam) = (\tilde{k}_1 \sigma_1 + \tilde{k}_2 \sigma_2 + \tilde{k}_3 \sigma_3)^2
		= \tilde{k}_1^2 + \tilde{k}_2^2 + \tilde{k}_3^2 \equiv \tilde{k}^2 .
\end{align}
By adding the following two terms,
\begin{align}
	(\bk \cdot \bgam) \gamma_i &= (\tilde{k}_1 \sigma_1 + \tilde{k}_2 \sigma_2 + \tilde{k}_3 \sigma_3)
	\nonumber\\
	&\times
	\left\{ W_i (1) \sigma_1 + W_i (2) \sigma_2 + W_i (3) \sigma_3 \right\}
	\nonumber\\
	&=
	\sum_\mu \tilde{k}_\mu W_i (\mu) + \Im \sum_{\lambda \mu \nu} \epsilon_{\lambda \mu \nu}
	\sigma_\lambda \tilde{k}_\mu W_i (\nu)
	\\
	\gamma_i (\bk \cdot \bgam)  &=\sum_\mu \tilde{k}_\mu W_i (\mu) - \Im \sum_{\lambda \mu \nu} \epsilon_{\lambda \mu \nu}
	\sigma_\lambda \tilde{k}_\mu W_i (\nu),
\end{align}
we obtain the commutation relation
	\begin{align}
		(\bk \cdot \bgam) \gamma_i + \gamma_i (\bk \cdot \bgam) =2\sum_\mu \tilde{k}_\mu W_i (\mu) \equiv 2R_i .
\end{align}
Also we have another commutation relation as
\begin{align}
	\gamma_i \gamma_j + \gamma_j \gamma_i &= 2\D \alpha_{ij},
\end{align}
since
\begin{align}
	\gamma_i \gamma_j &= 
	\left\{ W_i (1) \sigma_1 + W_i (2) \sigma_2 + W_i (3) \sigma_3 \right\}
	\nonumber\\
	&\times
	\left\{ W_j (1) \sigma_1 + W_j (2) \sigma_2 + W_j (3) \sigma_3 \right\}
	\nonumber\\
	&=
	\sum_\mu W_i (\mu) W_j (\mu) + \Im \sum_{\lambda \mu \nu} \epsilon_{\lambda \mu \nu}
	\sigma_\lambda W_i (\mu) W_j (\nu).
\end{align}

The followings are the trace for each term.

\underline{For $F_{\rm I}$:}
\begin{align}
	{\rm Tr} &\left[ (\bk \cdot \bgam) \gamma_y (\bk \cdot \bgam) \gamma_x (\bk \cdot \bgam) \gamma_y (\bk \cdot \bgam) \gamma_x \right]
	\nonumber\\
	&=
	{\rm Tr}\left[
	(\bk \cdot \bgam) \left\{ 2R_y - (\bk \cdot \bgam) \gamma_y \right\} \gamma_x
	(\bk \cdot \bgam) \left\{ 2R_y - (\bk \cdot \bgam) \gamma_y \right\} \gamma_x
	\right]
	\nonumber\\
	&=
	{\rm Tr}\Bigl[
	4R_y^2 (\bk \cdot \bgam) \gamma_x (\bk \cdot \bgam) \gamma_x
	-2R_y (\bk \cdot \bgam)(\bk \cdot \bgam)\gamma_y \gamma_x (\bk \cdot \bgam)\gamma_x
	\nonumber\\&
	-2R_y (\bk \cdot \bgam)\gamma_x(\bk \cdot \bgam)(\bk \cdot \bgam)\gamma_y \gamma_x 
	\nonumber\\
	&
	+(\bk \cdot \bgam)(\bk \cdot \bgam)\gamma_y \gamma_x(\bk \cdot \bgam)(\bk \cdot \bgam)\gamma_y \gamma_x
	\Bigr]
	\nonumber\\
	&=
	{\rm Tr}\Bigl[
	4R_y^2 (\bk \cdot \bgam) \left\{ 2R_x - (\bk \cdot \bgam)\gamma_x \right\} \gamma_x
	-2R_y \tilde{k}^2 \gamma_y \gamma_x (\bk \cdot \bgam) \gamma_x
	\nonumber\\&
	-2R_y \tilde{k}^2  (\bk \cdot \bgam) \gamma_x \gamma_y \gamma_x
	+\tilde{k}^4 \gamma_y \gamma_x \gamma_y \gamma_x
	\Bigr]
	\nonumber\\
	&=
	{\rm Tr} \Bigl[
	8 Rx R_y^2 (\bk \cdot \bgam) \gamma_x -4\D \alpha_{xx} R_y^2 \tilde{k}^2
	-8\D \alpha_{xy}R_y \tilde{k}^2 (\bk \cdot \bgam) \gamma_x 
	\nonumber\\&
	+ 4\D \alpha_{xx} R_y \tilde{k}^2 (\bk \cdot \bgam)\gamma_y
	\nonumber\\
	&+ \tilde{k}^4 (2\D \alpha_{xy}\gamma_y \gamma_x -\D^2 \alpha_{xx}\alpha_{yy})
	\Bigr]
	\nonumber\\
	&=
	2\left[
	8 R_x^2 R_y^2 -8\D \alpha_{xy} \tilde{k}^2 R_x R_y + \D^2 \tilde{k}^4 (2\alpha_{xy}^2 -\alpha_{xx}\alpha_{yy})
	\right]
\end{align}

\underline{For $F_{\rm II}$:}
\begin{align}
	{\rm Tr} &\left[ 2(\bk \cdot \bgam) \gamma_y (\bk \cdot \bgam) \gamma_x \gamma_y \gamma_x \right]
	\nonumber\\
	&=
	{\rm Tr} \left[
	2(\bk \cdot \bgam) \left\{ 2R_y -(\bk \cdot \bgam) \gamma_y \right\} \gamma_x \gamma_y \gamma_x \right]
	\nonumber\\
	&=
	{\rm Tr}\left[
	4 R_y (\bk \cdot \bgam) \gamma_x (2\D \alpha_{xy}-\gamma_x \gamma_y )-2\tilde{k}^2 \gamma_y \gamma_x \gamma_y \gamma_x
	\right]
	\nonumber\\
	&=
	2\left[
	8\D \alpha_{xy} R_x R_y - 4\D \alpha_{xx} R_y^2 - 2\D^2 \tilde{k}^2 (2\alpha_{xy}^2 - \alpha_{xx}\alpha_{yy})
	\right]
\end{align}

\underline{For $F_{\rm III}$:}
\begin{align}
	{\rm Tr} &\left[ 2(\bk \cdot \bgam) \gamma_x (\bk \cdot \bgam) \gamma_y \gamma_x \gamma_y \right]
	\nonumber\\
	&=
	{\rm Tr} \left[
	2(\bk \cdot \bgam) \left\{ 2R_x -(\bk \cdot \bgam) \gamma_x \right\} \gamma_y \gamma_x \gamma_y \right]
	\nonumber\\
	&=
	{\rm Tr}\left[
	4 R_x (\bk \cdot \bgam) \gamma_y (2\D \alpha_{xy}-\gamma_y \gamma_x )-2\tilde{k}^2 \gamma_x \gamma_y \gamma_x \gamma_y
	\right]
	\nonumber\\
	&=
	2\left[
	8\D \alpha_{xy} R_x R_y - 4\D \alpha_{yy} R_x^2 - 2\D^2 \tilde{k}^2 (2\alpha_{xy}^2 - \alpha_{xx}\alpha_{yy})
	\right]
\end{align}

\underline{For $F_{\rm IV}$:}
\begin{align}
	{\rm Tr}&\left[ (\bk \cdot \bgam) \gamma_y \gamma_x (\bk \cdot \bgam) \gamma_y \gamma_x \right]
	\nonumber\\
	&={\rm Tr} \left[
	(\bk \cdot \bgam) \gamma_y \left\{ 2R_x -(\bk \cdot \bgam)\gamma_x \right\} \gamma_y \gamma_x
	\right]
	\nonumber\\
	&=
	{\rm Tr}\left[
	2\D \alpha_{yy} R_x (\bk \cdot \bgam) \gamma_x - (\bk \cdot \bgam) \gamma_y (\bk \cdot \bgam) \gamma_x \gamma_y \gamma_x
	\right]
	\nonumber\\
	&=4\D \alpha_{yy}R_x^2 - \frac{1}{2}{\rm Tr}[F_{\rm II}]
\end{align}

\underline{For $F_{\rm V}$:}
\begin{align}
	{\rm Tr}&\left[ (\bk \cdot \bgam) \gamma_x  \gamma_y (\bk \cdot \bgam) \gamma_x \gamma_y \right]
	\nonumber\\
	&={\rm Tr} \left[
	(\bk \cdot \bgam) \gamma_x \left\{ 2R_y -(\bk \cdot \bgam)\gamma_y \right\} \gamma_x \gamma_y
	\right]
	\nonumber\\
	&=
	{\rm Tr}\left[
	2\D \alpha_{xx} R_y (\bk \cdot \bgam) \gamma_y - (\bk \cdot \bgam) \gamma_x (\bk \cdot \bgam) \gamma_y \gamma_x \gamma_y
	\right]
	\nonumber\\
	&=4\D \alpha_{xx}R_y^2 - \frac{1}{2} {\rm Tr}[F_{\rm III}]
\end{align}

\underline{For $F_{\rm II}+F_{\rm III}+F_{\rm IV}+F_{\rm V}$:}
\begin{align}
	F_{\rm II}+F_{\rm III}+F_{\rm IV}+F_{\rm V}&= \frac{1}{2}\left( F_{\rm II} + F_{\rm III}\right)
	+4\hbar^2\D \left( \alpha_{yy} R_x^2 + \alpha_{xx} R_y^2 \right)
	\nonumber\\
	&=
	4\hbar^2\D \left[ 4\alpha_{xy}R_x R_y - \D \tilde{k}^2 (2\alpha_{xy}^2 - \alpha_{xx}\alpha_{yy}) \right]
\end{align}

\underline{For $F_{\rm VI}$:}
\begin{align}
	{\rm Tr}\left[ \gamma_y \gamma_x \gamma_y \gamma_x \right]
	&=
	{\rm Tr}\left[
	\gamma_y \gamma_x (2\D \alpha_{xy} - \gamma_x \gamma_y )
	\right]
	\nonumber\\
	&=
	{\rm Tr} \left[
	2\D \alpha_{xy} \gamma_y \gamma_x - \D^2 \alpha_{xx}\alpha_{yy}
	\right]
	\nonumber\\&
	=
	2\D^2 (2 \alpha_{xy}^2 - \alpha_{xx}\alpha_{yy})
\end{align}


The final form of the trace for the matrix elements is
\begin{align}
	{\rm Tr} &\left[ \scr{G}\gamma_y \scr{G}\gamma_x \scr{G}\gamma_y \scr{G}\gamma_x \right]
	\nonumber\\
	&=
	2{\rm Tr}\left[
	\hbar^4 F_{\rm I} + \hbar^2 (\epsilon^2 -\D^2) \left( F_{\rm II} + F_{\rm III} + F_{\rm IV} + F_{\rm V}\right)
	+(\epsilon^2 -\D^2 )^2 F_{\rm VI}
	\right]
	\nonumber\\
	&=
	2{\rm Tr}\left[
	\hbar^4 F_{\rm I} + \hbar^2 (\epsilon^2 - E^2 + \hbar^2 \tk^2 )\left( F_{\rm II-V}\right)
	+(\epsilon^2 -E^2 + \hbar^2 \tk^2 )^2 F_{\rm VI}
	\right]
	\nonumber\\
	&=
	2{\rm Tr}\Bigl[
	(\epsilon^2 -E^2)^2 F_{\rm VI} + (\epsilon^2 -E^2) \left\{ 2\hbar^2 \tk^2 F_{\rm VI}+ F_{\rm II-V} \right\}
	\nonumber\\&
	+\hbar^4 \tk^4 F_{\rm VI} + \hbar^2 \tk^2 F_{\rm II-V} + F_{\rm I}
	\Bigr] ,
\end{align}
	where we used the relation $E^2 = \D^2 + \hbar^2 \tk^2$ and $\epsilon = \Im \tilde{\ve}_n$.

\section{Two-dimensional anisotropic Dirac electrons}\label{App_2D}
	
	We consider here a toy-model where both the electronic energy dispersion and the spin space are two-dimensional as in graphene with pseudo spins.\cite{Slonczewski1958,Ando2005}
	%
	%

\subsection{Spin Hall conductivity in two dimensions}

	The form of $\Phi_{{\rm s}jk}^i(\Im \omega_\lambda)$ for two-dimensions is the same as eq. (\ref{eq34}), except for the Jacobian $(\D^2 \det \hat{\alpha})^{-1/2}$, as 
\begin{align}
	\Phi_{{\rm s}jk}^i(\Im \omega_\lambda)
	&=-e\frac{\epsilon_{kji}\left( \alpha_{jk}^2 - \alpha_{jj}\alpha_{kk}\right)}{\sqrt{\D^2 \det \hat{\alpha}}}
	\nonumber\\
	&\times
	k_{\rm B}T\sum_{n\tk_\mu}
	\frac{4im\D^2 (\Im \tve_n - \Im \tve_{n-})}
	{\left\{ (\Im \tve_n)^2 -E_{\tk}^2\right\}\left\{ (\Im \tve_{n-})^2 -E_{\tk}^2\right\}}.
\end{align}
	The analytic continuation yields:
\begin{align}
	k_{\rm B}T&\sum_{n\tk_\mu}
	\frac{(\Im \tve_n - \Im \tve_{n-})}
	{\left\{ (\Im \tve_n)^2 -E_{\tk}^2\right\}\left\{ (\Im \tve_{n-})^2 -E_{\tk}^2\right\}}
	\nonumber\\
	&=-\frac{\hbar \omega}{2\pi \Im }\int_{-\infty}^{\infty}\!\! d\ve f(\ve) \sum_{\tk_\mu}
	\left[
	\frac{1}{\left(\ve_+^2 - E_{\tk}^2 \right)^2}-\frac{1}{\left(\ve_-^2 - E_{\tk}^2 \right)^2}
	\right]
	\nonumber\\
	&-\frac{\hbar \omega}{2\pi \Im }\int_{-\infty}^{\infty}\!\! d\ve \frac{d f(\ve)}{d\ve} \sum_{\tk_\mu}
	\frac{2i\Gamma}{\left(\ve_+^2 - E_{\tk}^2 \right)\left(\ve_-^2 - E_{\tk}^2 \right)}.
\end{align}
	The first term is the contribution from $C_1 + C_4$, and the second one is that from $C_2 + C_3$.
	Here we consider only the insulating case, so we discard the second term.
	The integration with respect to $\tk_{1, 2}$ is carried out as
	\begin{align}
		\sum_{\tk}\frac{1}{\left(\ve_+^2 - E_{\tk}^2 \right)^2}
		&=\frac{1}{2\pi} \int_0^{\infty} \!\! d\tk \, \tk
		\frac{1}{\left( \ve_+^2 - \D^2 -\hbar^2 \tk^2\right)^2}
		\nonumber\\
		&=-\frac{1}{4\pi \hbar^2}\frac{1}{\ve_+^2 -\D^2}.
\end{align}
	Then we obtain the two-dimensional spin-Hall conductivity for the insulating region as
	\begin{align}
	\sigma^i_{{\rm s}jk}(\Im \omega_\lambda)
	&=
	-\frac{me\D}{2\pi^2 \hbar \Im}
	\frac{\epsilon_{kji}\left( \alpha_{jk}^2 - \alpha_{jj}\alpha_{kk}\right)}{\sqrt{\det \hat{\alpha}}}
	K_{\rm 2D}^{\rm II},
	\end{align}
	where
	\begin{align}
		K_{\rm 2D}^{\rm II}=\int_{-\infty}^{\infty}\!\! d\ve f(\ve) 
	\left(
	\frac{1}{\ve_+^2 - \D^2}-\frac{1}{\ve_-^2 - \D^2}
	\right).
\end{align}

\subsection{Orbital susceptibility in two dimensions}
	
	The functional form of $M(\bk, \Im \tve_n)$ in two-dimensions is the same as eq. (\ref{eq63}):
	\begin{align}
	&{\rm Tr}[M(\bk, \Im \tve_n)]
	\nonumber\\
	&=-4\D^2 \alpha_{xx}\alpha_{yy}\left\{ (\Im \tve_n)^2 -E_{\tk}^2 \right\}^2
	+8\left[ \D \alpha_{xy}\left\{ (\Im \tve_n)^2 -E_{\tk}^2 \right\}
	+2\hbar^2 R_x R_y \right]^2.
\end{align}
	The contributions from $\bm{R}$ are
	\begin{align}
	R_x R_y &= \pi \D\tk \alpha_xy,\\
	R_x^2 R_y^2 &=\frac{\pi}{4}\D^2 \tk^4 (2\alpha_{xy}^2 + \alpha_{xx}\alpha_{yy}).
	\end{align}
	After the integration with respect to the angle, we obtain
	\begin{align}
		{\rm Tr}[M(\bk, \Im \tve_n)]
		&=8\pi \D^2 \Bigl[
		(2\alpha_{xy}^2- \alpha_{xx}\alpha_{yy})\left\{ (\Im \tve_n)^2 -E_{\tk}^2 \right\}^2
		\nonumber\\&
	+4 \alpha_{xy}^2 \hbar^2 \tk^2 \left\{ (\Im \tve_n)^2 -E_{\tk}^2 \right\}
	+(2\alpha_{xy}^2 + \alpha_{xx}\alpha_{yy})\hbar^4 \tk^4
	\Bigr].
\end{align}
	The integrations associated with the Green's function in two-dimensions are as follows:
	\begin{align}
		k_{\rm B}T&\sum_n \sum_{\tk_{1,2}}\frac{1}{\left[ (\Im \tve_n)^2 -E_{\tk}^2 \right]^2}
		\nonumber\\
		&=
		\frac{1}{8\pi^2 \hbar^2 \Im}
		\int_{-\infty}^{\infty}\!\! d\ve f(\ve) 
	\left(
	\frac{1}{\ve_+^2 - \D^2}-\frac{1}{\ve_-^2 - \D^2}
	\right),
	\\
	k_{\rm B}T&\sum_n \sum_{\tk_{1,2}}\frac{\hbar^2 \tk^2}{\left[ (\Im \tve_n)^2 -E_{\tk}^2 \right]^3}
		\nonumber\\
		&=
		-\frac{1}{16\pi^2 \hbar^2 \Im}
		\int_{-\infty}^{\infty}\!\! d\ve f(\ve) 
	\left(
	\frac{1}{\ve_+^2 - \D^2}-\frac{1}{\ve_-^2 - \D^2}
	\right),
	\\
	k_{\rm B}T&\sum_n \sum_{\tk_{1,2}}\frac{\hbar^4 \tk^4}{\left[ (\Im \tve_n)^2 -E_{\tk}^2 \right]^4}
		\nonumber\\
		&=
		\frac{1}{24\pi^2 \hbar^2 \Im}
		\int_{-\infty}^{\infty}\!\! d\ve f(\ve) 
	\left(
	\frac{1}{\ve_+^2 - \D^2}-\frac{1}{\ve_-^2 - \D^2}
	\right).
\end{align}
	Then we have
	\begin{align}
	k_{\rm B}T&\sum_{n, \tk_{1,2}}{\rm Tr}\left[ \scr{G}v_y\scr{G}v_x\scr{G}v_y\scr{G}v_x \right]
	\nonumber\\
	&= \frac{\D^2}{2\pi^2 \hbar^2}
	\left[ (2\alpha_{xy}^2 -\alpha_{xx}\alpha_{yy})-2 \alpha_{xy}^2 + \frac{1}{3}(2\alpha_{xy}^2 + \alpha_{xx}\alpha_{yy})
	\right] K_{2D}^{\rm II}
	\nonumber\\
	&=
	\frac{\D^2}{2\pi^2 \hbar^2 \Im}
	\frac{2}{3}(\alpha_{xy}^2 -\alpha_{xx}\alpha_{yy})K_{\rm 2D}^{\rm II}.
\end{align}
	Finally, we obtain the two-dimensional orbital susceptibility in the form
	\begin{align}
		\chi^z =\frac{e^2 \D}{6\pi^2 c^2 \Im }\frac{(\alpha_{xy}^2 -\alpha_{xx}\alpha_{yy})}{\sqrt{\det \hat{\alpha}}}K_{\rm 2D}^{\rm II}.
\end{align}

	The ratio of the two-dimensional SHC and orbital susceptibility in the insulating state is
	\begin{align}
		\frac{\sigma_{{\rm s}xy}^z}{\chi^z}
		=\frac{me\D }{2\pi^2 \hbar \Im}\frac{6\pi^2 c^2 \Im}{e^2 \D}
		=\frac{3mc^2}{\hbar e}.
\end{align}
	Namely, the relation eq. (\ref{relation}) still holds even for the two-dimensional systems, though the functional forms of $\sigma_{{\rm s}xy}$ and $\chi$ are different from those for the three-dimensional systems.

\subsection{Functional form of $K_{\rm 2D}^{\rm II}$}
	The function $K_{\rm 2D}^{\rm II}$ can be rewritten as
	\begin{align}
		K_{\rm 2D}^{\rm II} 
	&=\int_{-\infty}^{\infty}\!\! d\ve f(\ve) \frac{1}{2\D}
	\left[
	\frac{-2\Im \Gamma}{(\ve-\D)^2+\Gamma^2} -\frac{-2\Im \Gamma}{(\ve+\D)^2+\Gamma^2}
	\right].
	\end{align}
	In the $\Gamma \to 0$ limit at zero temperature, we have
	\begin{align}
	K_{\rm 2D}^{\rm II}&= \frac{\pi \Im }{\D}\int_{-\infty}^{\infty}\!\! d\ve f(\ve) 
	\left[ \delta(\ve+\D) - \delta(\ve-\D) \right]
	=\frac{\pi \Im }{\D}\theta (\D - |\mu|),
\end{align}
	where $\theta (x)$ is the step function.
	Consequently, the two-dimensional orbital susceptibility becomes
	\begin{align}
		\chi^z =\frac{e^2 }{6\pi c^2 }\frac{(\alpha_{xy}^2 -\alpha_{xx}\alpha_{yy})}{\sqrt{\det \hat{\alpha}}}\theta (\D - |\mu|),
\end{align}
	i.e., $\chi$ is finite only in the insulating state.
	In the isotropic case, $\alpha_{\mu \mu}=\gamma^2/\D$, it is given by
	\begin{align}
		\chi^z =-\frac{e^2 \gamma^2}{6\pi c^2 \D}\theta (\D - |\mu|),
\end{align}
	which agrees with the previous results\cite{Koshino2010}.

\bibliographystyle{jpsj}
\bibliography{Bismuth,footnote}

\begin{thebibliography}{10}
\bibitem{note2}
When we carry out the integration first with respect to $\ve$, and then to
  $\bk$, the main contributions are the same as eqs. (\ref{K1clean}) and
  (\ref{K2clean}), though a small constant term $+1$ appears in addition to
  them.
\bibitem{note3}
The SHE in the metallic state is accompanied by the Joule heating due to the
  electric current. On the other hand, for the SHE in insulating state, the
  electric current does not flow, so that the spin-current can flow without
  Joule heating.
\end{thebibliography}
\end{document}